# FISICA: The Florida Image Slicer for Infrared Cosmology & Astrophysics

Stephen Eikenberry, S. Nicholas Raines, Nicolas Gruel, Richard Elston, Rafael Guzman, Jeff Julian
Department of Astronomy, University of Florida, Gainesville, FL

Glenn Boreman
Center for Research in Electro-Optics & Lasers, University of Central Florida, Orlando, FL

Paul Glenn, Greg Hull-Allen
Bauer Associates, Wellesley, MA

Jeff Hoffmann, Michael Rodgers, Kevin Thompson
Optical Research Associates, Pasadena, CA

Scott Flint, Lovell Comstock, Bruce Myrick
Corning NetOptix, Keene, NH

## ABSTRACT

We report on the design, fabrication, and on-sky performance of the Florida Image Slicer for Infrared Cosmology and Astrophysics (FISICA) – a fully-cryogenic all-reflective image-slicing integral field unit for the FLAMINGOS near-infrared spectrograph.  Designed to accept input beams near f/15, FISICA with FLAMINGOS provides R~1300 spectra over a 16x33-arcsec field-of-view on the Cassegrain f/15 focus of the KPNO 4-meter telescope, or a 6x12-arcsec field-of-view on the Nasmyth or Bent Cassegrain foci of the Gran Telescopio Canarias 10.4-meter telescope.  FISICA accomplishes this using three sets of "monolithic" powered mirror arrays, each with 22 mirrored surfaces cut into a single piece of aluminum.  We review the optical and opto-mechanical design and fabrication of FISICA, as well as laboratory test results for FISICA integrated with the FLAMINGOS instrument.  Finally, we present performance results from observations with FISICA at the KPNO 4-m telescope and comparisons of FISICA performance to other available IFUs on 4-m to 8-m-class telescopes.

## 1. INTRODUCTION

Integral field capabilities have become a key contributor to many current spectroscopic research programs in optical and infrared astronomy.  While optical fibers have tended to work rather well in integral field units (IFU) for optical instruments, they can suffer from many drawbacks for infrared work, including low throughput, limited bandwidth, high background (for non-cryogenic IFUs), and other optical drawbacks as well.  An alternative approach is to use a segmented set of tilted mirrors at an image plane to "slice" the input 2-dimensional image and re-format it into a pseudo-longslit input to a spectrograph.  Excellent descriptions of designs for such systems are provided by Content (1997) and Content (1998).

Several previous image-slicing IFUs based on slicer mirrors have been built or are currently under construction (i.e. 3D – Krabbe et al., (1997); PIFS – Murphy et al., 1999; UIST IFU – Todd, et al., (2003); SPIFFI – Tecza et al., (2003); GNIRS IFU – Dubbledam et al., (2000); NIFS – McGregor et al. (2003)).  However, each of these has one or more significant drawbacks, such as: non-powered slicer mirrors; small fields of view (only a few square arcseconds on 8- to 10-meter telescopes); multi-element mirror arrays (requiring painstaking alignment and mounting of a large number of mechanically-fragile elements); heterogeneous materials (causing differential misalignment from room to cryogenic temperatures). FISICA avoids or mitigates each of these problems, using monolithically-fabricated, all-aluminum powered mirror arrays to provide a >70-square-arcsecond field on the Gran Telescopio Canarias 10-meter telescope, or a ~500-square-arcsecond field on the KPNO 4-meter telescope.

FISICA (Eikenberry et al., 2004) is a 22-slice advanced slicer IFU design, incorporating three powered mirror arrays of 22 elements each at – the "slicer" mirror array, the "pupil" mirror array, and the "field" mirror array. It is designed to fit inside a clone of the multi-object spectroscopy (MOS) dewar of the FLAMINGOS instrument (Elston et al., 1998; Elston et al., 2002), slicing the input telescope focal plane and then replacing it in the outgoing beam such that the virtual sliced image appears at a virtual focus coincident with the input telescope focal plane. Thus, FLAMINGOS effectively "sees" a simple pseudo-longslit input coming as if from the telescope itself, requiring no modifications to the existing instrument.

In Section 2 below, we describe the optical concept for FISICA. In Section 3, we discuss the opto-mechanical design for the instrument, including the slicer mirror arrays. In Section 4, we present the fabricated results for FISICA, and in Section 5 we show recent laboratory test results from FISICA integrated with FLAMINOGS in the laboratory. Finally, in Section 6 we discuss our plans for the upcoming first-light observing runs with FISICA on the KPNO 4-meter telescope in July 2004.

## 2. FISICA OPTICAL CONCEPT

As noted above, the basic optical principle behind FISICA is to take an input 2-dimensional field-of-view at the telescope focal plane, divide it into 22 "slices", orient these slices end-to-end to create a pseudo-longslit, and then place a virtual image of this pseudo-longslit in the optical beam so that it appears to originate from the output telescope focal plane. The basic design we used to accomplish this task is shown in Figure 1.

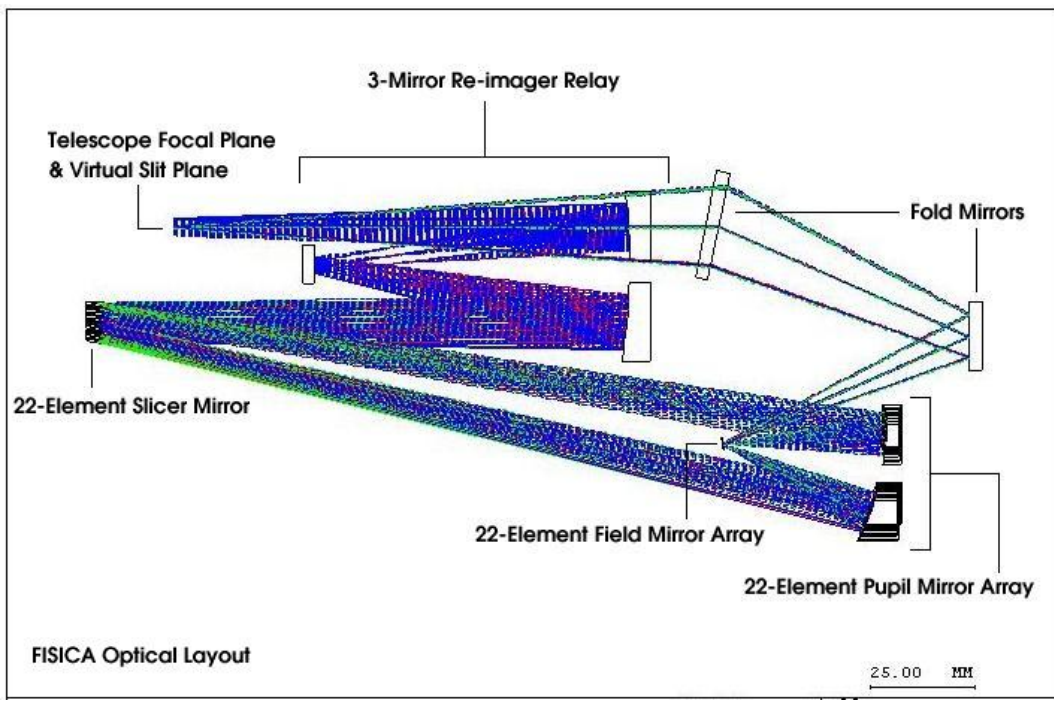

**Figure 1** – Optical concept layout for FISICA

In Figure 1, light from the telescope enters from the left and forms an image at the telescope focal plane. The expanding beam (up to f/15) encounters a 3-mirror re-imaging relay, which magnifies the image by x2, and comes to a focus at the surface of the 22-element "slicer" mirror. The slicer mirror has 22 slices, each powered, each tilted along both the "spatial" direction of the spectrograph (in order to separate the final images into a pseudo-longslit) and in the "dispersion" direction of the spectrograph (in order to separate pupil images into 2 rows of 11 each). The powered slicer mirror creates an image of the telescope exit pupil on the 22-element pupil mirror array. The pupil mirror array consists of 22 powered, offset, tilted mirrors arranged in two rows of 11 mirrors each. This 2x11 geometry was chosen in order to minimize field angles for the IFU, and thus reduce aberrations. The pupil mirrors create another relayed image of the telescope focal plane along a linear array of 22 field mirrors. This image is demagnified 4x from the slicer mirror, or 2x from the original input beam (thus matching the f/15 telescope beam to the f/8 spectrograph beam of FLAMINOGS). We chose this geometry to allow the pupil mirrors to be significantly larger than the closely-spaced field mirrors, and for the field mirror array to correct telecentricity errors, and thus "offload" some tilt from the pupil mirror array. Two fold mirrors then relay the output "sliced" image to the FLAMINGOS spectrograph. The actual final rays proceed out of the figure to the right from the last fold mirror, while the figure shows them "reverse-traced" to the left to show that they do indeed appear to come from the output of the telescope focal plane.

## 3. FISICA OPTO-MECHANICAL CONCEPT

The basic opto-mechanical concept for FISICA is to house the entire IFU in a clone of the multi-object spectroscopy (MOS) dewar of FLAMINGOS. With this arrangement, the FISICA dewar can be exchanged with the FLAMINGOS MOS dewar in a rapid daytime operation, switching the instrument from MOS to IFU mode from one night to the next. We show the opto-mechanical layout of FISICA in Figures 2. The FISICA optics are mounted in an aluminum reference structure which bolts to the dewar work surface, and primarily extends into the center of the toroidal nitrogen reservoir towards the spectrograph dewar.

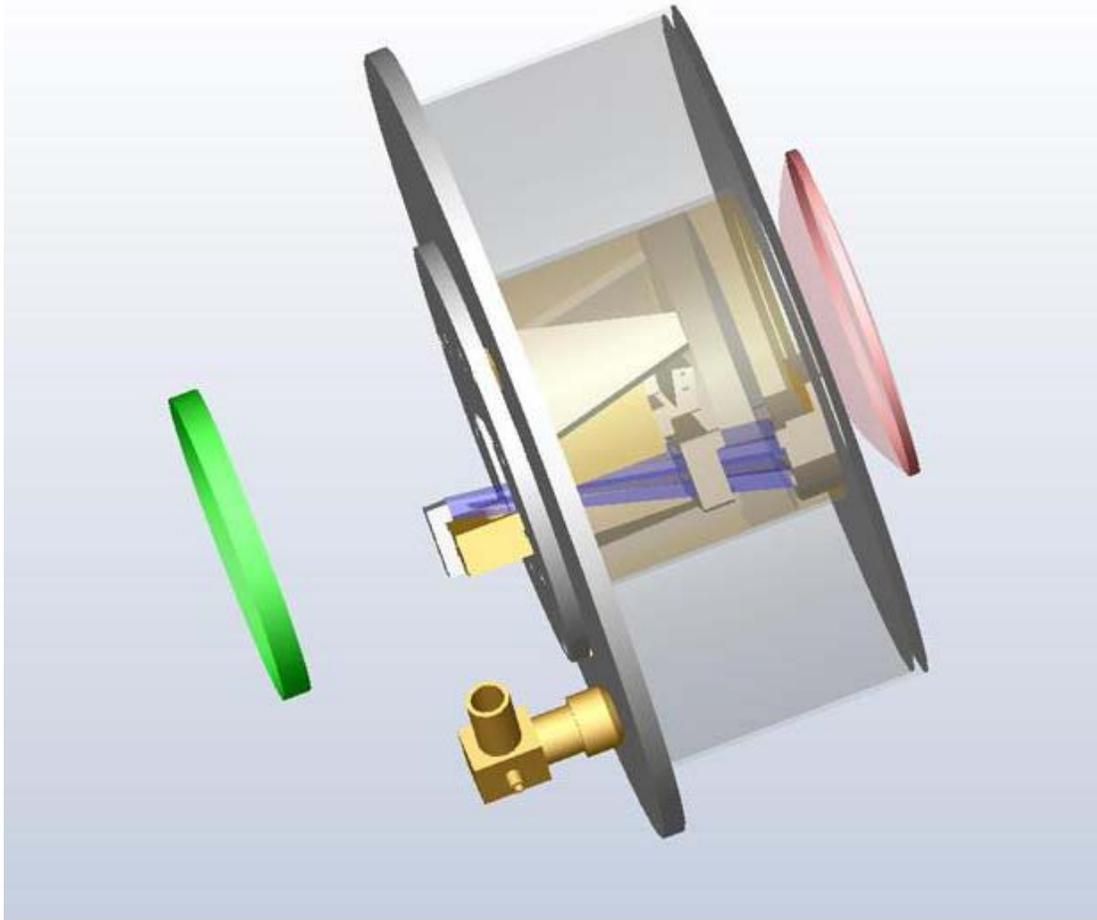

**Figure 2** – Conceptual layout of FISICA in a clone of the FLAMINGOS MOS dewar. The disk at the left is the FISICA dewar entrance window. The FISICA IFU assembly id located inside the torus of the dewar nitrogen reservoir. Light exits to the right, through the field lens which serves as the entrance window for the FLAMINGOS spectrograph dewar.

As noted above, FISICA has some significant differences from previous image-slicing IFU instruments.  Most prominently, the "slice" optics in three mirror arrays (slicer array, pupil array, and field mirror array) are manufactured from just 3 pieces of aluminum – one for each mirror array.  The details of the design strategy for this approach are detailed in a companion paper in these proceedings (Glenn et al., 2004).  However, the basic philosophy behind this is that monolithic mirror arrays are mechanically very robust and cryogenically reliable (as compared to clamping together 22 delicate individually-fabricated mirrors for each array), and they are MUCH easier to align.  Essentially, all between-slice alignments are offloaded to the manufacturing process, so that only 3 moving parts exist for all 66 "slice" mirrors.  Secondly, the mirrors are all diamond-turned from the same billet of 6061-T6 aluminum and the same alloy as the IFU structural elements.  As a consequence of this precise CTE-matching, the entire IFU contracts homologously, so that alignments done at room temperature will remain valid at cryogenic temperatures.

## 4. FISICA FABRICATION RESULTS

All component fabrication for FISICA was completed in early 2004.  In Figure 3, we show each of the three key mirror arrays for FISICA – the slicer mirror array, pupil mirror array, and field mirror array.

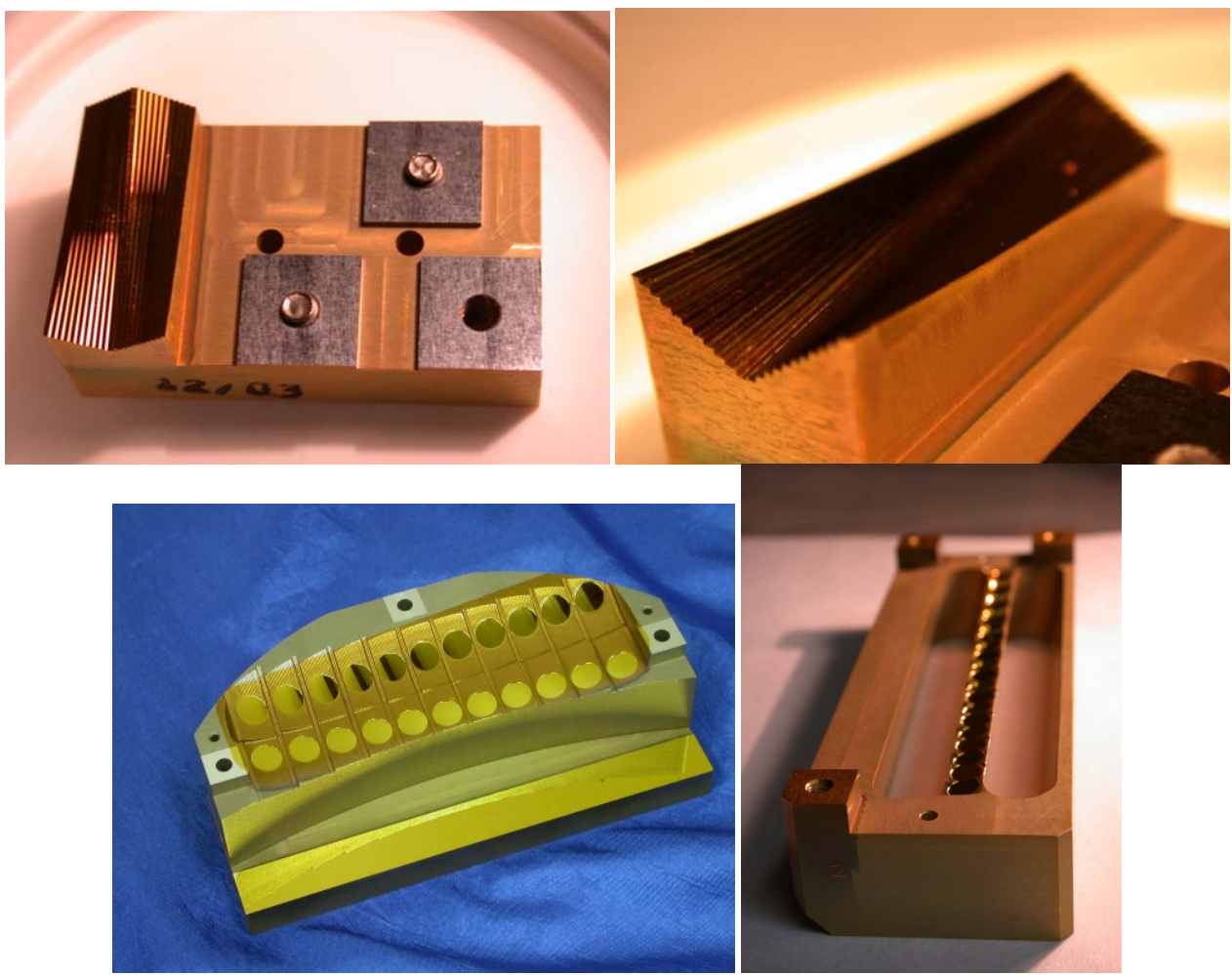

**Figure 3** – Fabricated mirror arrays for FISICA prior to integration with the IFU structure.  (Top left) monolithic slicer mirror array and base.  (Top right) close-up of the slicer array; (Bottom left) monolithic pupil mirror array with 2x11 geometry.  (Bottom right) monolithic field mirror array.

In Figure 4, we show the assembly of the IFU in its reference structure, and in Figure 5 we show the IFU in the FISICA dewar.

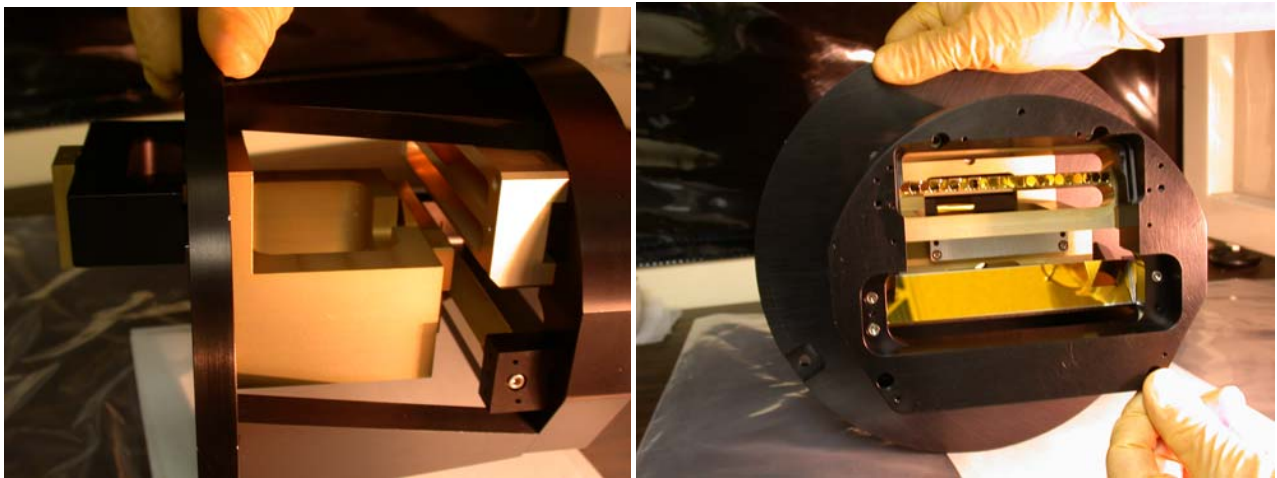

**Figure 4** – FISICA IFU assembly with all optical components integrated. (Left) side view. (Right) rear view. Note that the linear field mirror array is visible here near the top of the IFU (as seen via the fold mirrors).

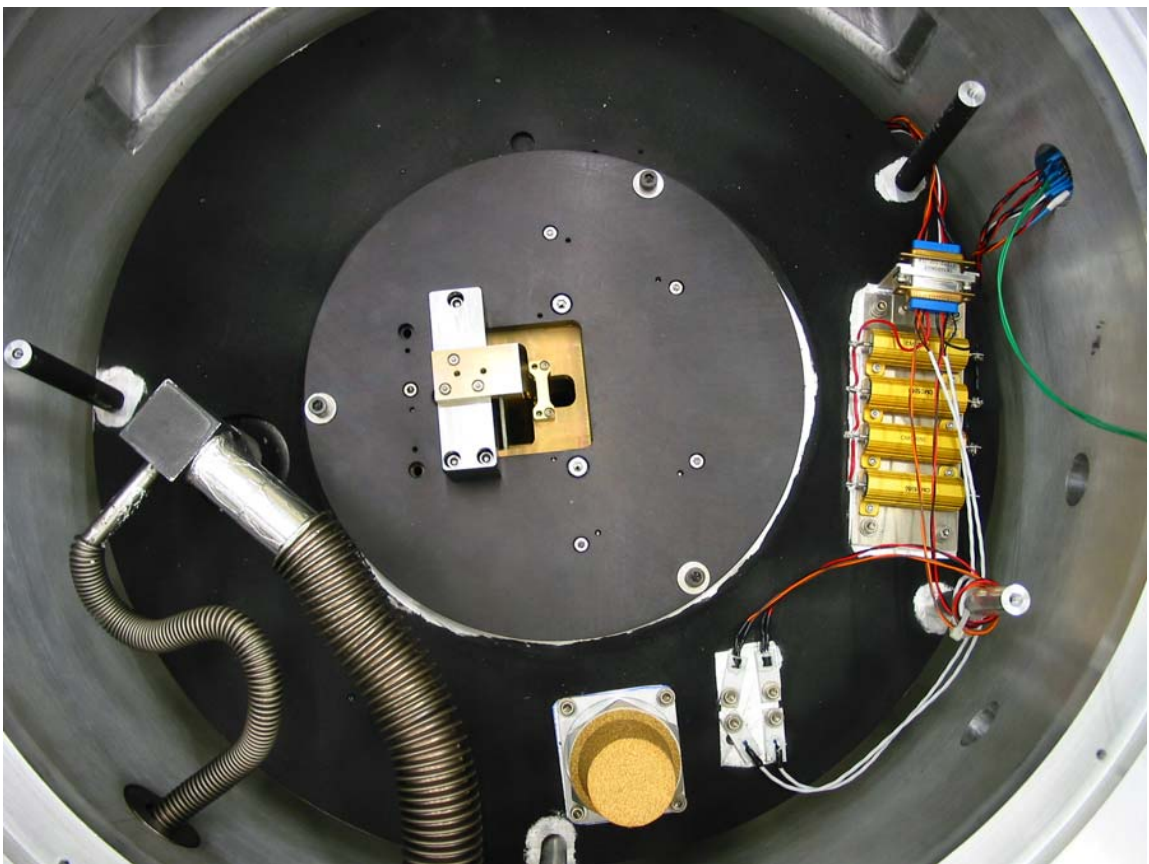

**Figure 5** – FISICA IFU assembly integrated inside the FISICA dewar.

## 5. FISICA PERFORMANCE

The detailed laboratory testing of the FISICA IFU itself is described in a companion paper (Glenn et al., 2004). In this section, we describe the laboratory test results of FISICA integrated with the FLAMINGOS spectrograph. Our primary test was to create an artifical "dome flat" on the ceiling of the laboratory, and use it to illuminate FISICA. We then used Hartmann masks in the FLAMINGOS Lyot wheel to verify the focus location of the IFU relative to the spectrograph. Figure 6 shows the pseudo-longslit output of the IFU as imaged by FISICA/FLAMINGOS.

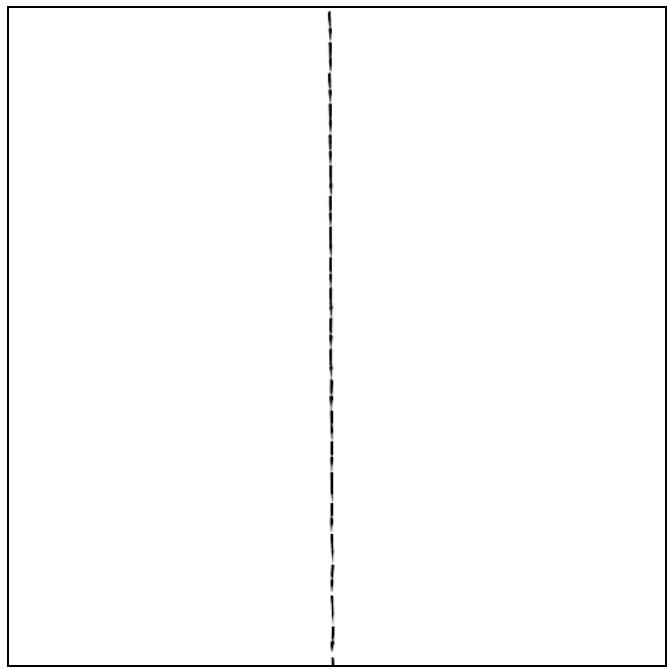

**Figure 6** – Sliced image of a laboratory "dome flat" through FISICA/FLAMINGOS, taken in the H-band at cryogenic temperature. The slices are clearly visible, and as expected form a pseudo-longslit.

One problem noted with the slit images, as described in Glenn et al. (2004), are "dimples" in the centers of the field mirrors. These cause a near-complete loss of light over ~1-arcsec near the center of each affected slitlet (comprising ½ of all slitlets). While these were somewhat problematic, they can be compensated using standard dither/nod observing techniques. Furthermore, by slightly offsetting the flat field mirror arrays, we succeeded in moving the slit images off the dimples. This mechanical adjustment was made after the first-light observations with FISICA in July 2004.

We commissioned FISICA in July 2004 on the KPNO 4-m telescope. We have had several additional observing runs since then, during which we have fully characterized FISICA's on-telescope performance. Figure 7 shows an image through a J-band filter (no grism) of the FISICA field and a software-reconstructed 2-D image of the field. The star in the field has a FWHM ~0.9-arcsec, limited by the seeing at that time. In 2005, we obtained images with FWHM ~0.7-arcsec (2 pixels), again limited by the atmospheric seeing as evidenced by the smooth, round nature of the PSFs. These observations confirm that FISICA does not significantly degrade atmospheric seeing even at the Nyquist-limited specification for the instrument. Figure 8 shows a false color reconstruction of the starburst galaxy NGC 1569, incorporating several spectral features detected in an early FISICA observation at KPNO on the 4-m telescope.

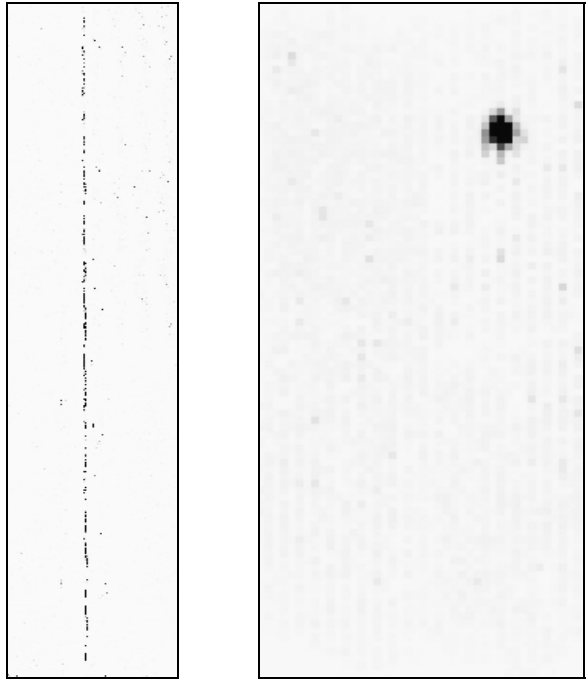

Figure 7 – (Left) "Through-IFU" image of field, as seen on the FISICA HAWAII-2 detector. (Right) Software-reconstructed 2-D field map from the same image. The star image has FWHM 0.9-arcsec in this J-band image.

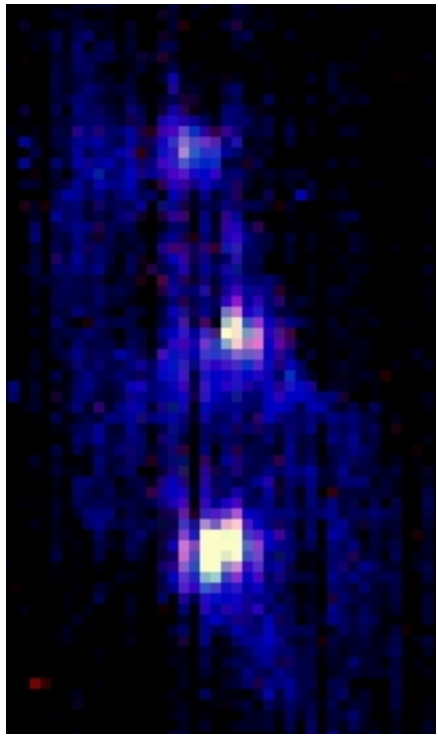

Figure 8 – FISICA false color "spectral image of the dwarf starburst galaxy NGC1569. Red is Paβ, green is H-band continuum, blue is HeI (1.083 μm). Note that the HeI emission extends well beyond the Paβ emission, which is concentrated in the previously known superstarclusters.

In Table 1, we compare the performance of the FISICA IFU to several other IFUs in operation. Generally speaking, FISICA significantly outperforms the competition – even those on much larger 8-m telescopes at better-seeing sites – especially for surveys of extended regions (similar to or larger than the FISICA field of view). In the cases of UIST and CIRPASS, this is due in part to the significantly higher optical throughput of FISICA. In all cases, FISICA offers a significantly larger "$A\Omega$" product than the other IFUs.

Table 1 – Comparison of FISICA to Other IFUs

|                        | Figure of Merit Ratio |                  |                |
| ---------------------- | --------------------- | ---------------- | -------------- |
| **Figure of Merit**    | **FISICA/UIST-IFU**   | **FISICA/GNIRS-IFU** | **FISICA/CIRPASS** |
| $A\Omega$              | 16.7                  | 7.2              | 3.3            |
| $A\Omega/\sigma^2$     | 4.2                   | 1.8              | 0.8            |
| $\eta\, A\Omega/\sigma^2$ | 6.1                | 1.9              | 3.1            |

The figure-of-merit variables used in the above are telescope collecting area (A), IFU areal field-of-view ($\Omega$), seeing-limited PSF width ($\sigma$), and IFU throughput ($\eta$).

The authors acknowledge the support of the University of Florida, the University of Central Florida, the Florida Space Research Initiative, and the Fundacion por la Preservacion del PSF.